\documentclass[useAMS,usenatbib]{mnras}

\usepackage{amssymb,graphicx,times}
\usepackage{booktabs}
\usepackage{multirow}
\usepackage{aas_macros}
\usepackage{hyperref}
\usepackage{breakurl}

\pagerange{\pageref{firstpage}--\pageref{lastpage}} \pubyear{2015}

\begin{document}

\label{firstpage}

\title[Radio/$\gamma$-ray cross-correlation]{Locating the $\gamma$-ray emission site in \textit{Fermi}/LAT blazars from correlation analysis between 37~GHz radio and $\gamma$-ray light curves}
\author[Ramakrishnan et~al.]{V.~Ramakrishnan,$^{1}$\thanks{E-mail: venkatessh.ramakrishnan@aalto.fi} T.~Hovatta,$^{1,2}$ E.~Nieppola,$^{1,3}$ M.~Tornikoski,$^{1}$ A.~L\"{a}hteenm\"{a}ki$^{1,4}$ 
	\newauthor
	and E.~Valtaoja$^{5}$ \\
	\\
$^{1}$Aalto University Mets\"{a}hovi Radio Observatory, Mets\"{a}hovintie 114, 02540, Kylm\"{a}l\"{a}, Finland\\
$^{2}$Owens Valley Radio Observatory, California Institute of Technology, Pasadena, CA 91125, USA\\
$^{3}$Finnish Centre for Astronomy with ESO (FINCA), University of Turku, V\"{a}is\"{a}l\"{a}ntie 20, 21500 Piikki\"{o}, Finland\\
$^{4}$Aalto University Department of Radio Science and Engineering, PL 13000, FI-00076 Aalto, Finland\\
$^{5}$Tuorla Observatory, Department of Physics and Astronomy, University of Turku, 20100 Turku, Finland}

\maketitle

\begin{abstract}
	We address the highly debated issue of constraining the $\gamma$-ray emission region in blazars from cross-correlation analysis using discrete correlation function between radio and $\gamma$-ray light curves. The significance of the correlations is evaluated using two different approaches: simulating light curves and mixed source correlations. The cross-correlation analysis yielded 26 sources with significant correlations. In most of the sources, the $\gamma$-ray peaks lead the radio with time lags in the range +20 and +690 days, whereas in sources 1633+382 and 3C~345 we find the radio emission to lead the $\gamma$ rays by -15 and -40 days, respectively. Apart from the individual source study, we stacked the correlations of all sources and also those based on sub-samples. The time lag from the stacked correlation is +80 days for the whole sample and the distance travelled by the emission region corresponds to 7~pc. We also compared the start times of activity in radio and $\gamma$ rays of the correlated flares using Bayesian block representation. This shows that most of the flares at both wavebands start at almost the same time, implying a co-spatial origin of the activity. The correlated sources show more flares and are brighter in both bands than the uncorrelated ones.

\end{abstract}


\section{Introduction}

Blazars (i.e., flat-spectrum radio quasars, FSRQs, and BL~Lac objects) are a type of active galactic nuclei (AGN) with a jet oriented close to the line of sight. The broadband spectral energy distribution of blazars is characterized by the presence of two broad bumps. The low energy component (from radio to UV or, in some cases, X-rays) is produced via synchrotron radiation by relativistic electrons in the jet plasma. The high energy component, generally peaking at the $\gamma$-ray regime, can be produced by inverse Compton (IC) scattering of the same electrons responsible for the synchrotron emission (leptonic models) or emission resulting from the cascades initiated by photo-pair and photo-pion production of ultrarelativistic hadrons in the jet (hadronic models). For further information on leptonic and hadronic blazar models, see \citet{Bottcher2013}.

Despite many studies, the knowledge on the location of the $\gamma$-ray emission and its mechanism remains uncertain. Several models have been proposed regarding the $\gamma$-ray emission site relative to the central engine in blazars. Some of them constrain the location closer to the supermassive black hole ($<0.1$--1~pc) based on the observed rapid $\gamma$-ray variability in few sources \citep[e.g.,][]{Tavecchio2010,Foschini2011}. In such cases, the $\gamma$-ray emission could be generated by IC process on the external photons (EC), in which the soft photons are directly from the accretion disc \citep[e.g.,][]{Dermer2002} or from the broad line region \citep[BLR; e.g.,][]{Sikora1994}.

Multifrequency and relative timing analysis of individual sources constrains the $\gamma$-ray emission site to be on parsec scales in the relativistic jets of blazars. Results from single-dish radio/mm observations shows that the $\gamma$-ray flares are typically found between the onset and the peak of the mm outburts \citep{AL2003,Jonathan2011}. Connection of $\gamma$-ray flares to the propagation of superluminal components in the jets observed using very long baseline interferometry (VLBI) suggests that the kinematics of shocks or shock--shock interaction at parsecs down the jet as possible sites for the $\gamma$-ray emission \citep[e.g.,][]{Jorstad2001,Jorstad2013,Agudo2011a,Agudo2011b}. Under this scenario, the $\gamma$-ray emission could either be generated via IC scattering of seed photons that originate as synchrotron emission in the jet \citep[synchrotron self-Compton or SSC; e.g.,][]{Bloom1996} or from the jet consisting of highly-relativistic spine surrounded by a mildly-relativistic sheath as a source of EC \citep[spine-sheath model;][]{Marscher2010}.

Thus, one way to address this question is by multifrequency correlation studies that can constrain the location of emission region. Recent works such as, \citet{Cohen2014}, \citet{Fuhrmann2014} and \citet{MaxMoerbeck2014a}, investigate the correlations between $\gamma$ rays and optical and radio light curves. These works have utilised the $\gamma$-ray data provided by the ongoing successful mission of \textit{Fermi Gamma-ray Space Telescope} along with its primary scientific instrument the Large Area Telescope (\textit{Fermi}/LAT) providing unprecedented coverage of the $\gamma$-ray sky.

In this work, we discuss the results from the correlation analysis between the 37~GHz radio light curves obtained from the Mets\"{a}hovi AGN monitoring programme and the \textit{Fermi} $\gamma$-ray light curves of 55 blazars. The paper is organised as follows. In Section~\ref{sample_def} we describe the sample definition used for selecting 55 blazars. We describe the observation and data reduction methods in Section~\ref{obs_datred}. Sections~\ref{dcf_method} and \ref{dcf_sign} describe the correlation and various significance methods used in this work. The results and discussion of the correlations are presented in Sections~\ref{results} and \ref{discuss} before summarising the work in Section~\ref{conclude}.

We assume a $\Lambda$ cold dark matter cosmology throughout this work with $H_{\rm 0}=68~{\rm km~s^{-1}Mpc^{-1}, \Omega_m=0.3}$, and $\Omega_{\Lambda}=0.7$ \citep{Planck2014}. The term radio core used in many places in this work refers to the compact unresolved feature in the VLBI maps that is usually located at one end of the jet.

\section{Sample definition}
\label{sample_def}

For this work, we considered all AGNs from the Mets\"{a}hovi monitoring programme with observations since 2008.6 that are also part of the second \textit{Fermi} Gamma-ray catalogue \citep[hereafter 2FGL; ][]{Nolan2012}. We required the 37~GHz mean flux density to be above 1~Jy for the time period 2008.6--2013.6. We obtained 55 AGNs which were also variable at the $3\sigma$ level according to a $\chi^2$ test. Our sample could be sub-divided into 40 FSRQs, 14 BL~Lacs and 1 radio galaxy. The source names, optical classification and redshifts are given in Table~\ref{tab1}.

\section{Observations and Data Reduction}
\label{obs_datred}

\subsection{\textit{Fermi}/LAT}
The \textit{Fermi}/LAT is an electron-positron pair conversion telescope covering the energy range from 20~MeV to $>$300~GeV \citep{Atwood2009}. It operates in a survey mode observing the entire sky every 3 hr. This makes \textit{Fermi} an ideal source of $\gamma$-ray data for our analysis. 

The $\gamma$-ray fluxes for the energy range 0.1--200~GeV were obtained by analysing the \textit{Fermi}/LAT data from 2008 August 4 to 2013 August 3 (encompassing 5~yr of $\gamma$-ray data) using the Fermi Science Tools\footnote{\url{http://fermi.gsfc.nasa.gov/ssc/data/analysis/documentation/Cicerone}} version v9r32p5. Following the data selection recommendations\footnote{\url{http://fermi.gsfc.nasa.gov/ssc/data/analysis/documentation/Cicerone/Cicerone_Data_Exploration/Data_preparation.html}}, we select photons in event class 2, with a further selection on the zenith angle $>100^{\circ}$ to avoid contamination from the photons coming from the Earth's limb. The photons were extracted from a circular region centered on the source, within a radius of 15$^{\circ}$. The instrument response functions P7REP\_SOURCE\_V15 were used.

We implemented an unbinned likelihood analysis using {\it gtlike} \citep{Cash1979, Mattox1996}, modelling all the sources within $15^{\circ}$ from our source position (region-of-interest, ROI) obtained from the 2FGL. We fixed the model parameters of sources with test statistic (TS) $<4$ to the 2FGL value and also for sources $>10^{\circ}$ from ROI. The normalization was left free, while the spectral indices for all the sources were fixed at the 2FGL value. The Galactic diffuse emission and the isotropic background (sum of extragalactic diffuse and residual instrumental backgrounds) were also modelled at this stage using the templates 'gll\_iem\_v05.fits' and 'iso\_source\_v05.txt' provided with the Science Tools. Our final fluxes were obtained from 7~d integrations with a detection criterion such that, the maximum-likelihood TS \citep{Mattox1996} exceeds four ($\sim 2\sigma$). Bins with TS $<4$ or when the predicted number of photons are less than four, $2\sigma$ upper limits were estimated using the profile likelihood method \citep{Rolke2005}.

In addition to the weekly binned light curve, we obtained a light curve with 30~d integrations using the unbinned likelihood analysis to account for the effect of sparse sampling of certain sources in our radio sample during the correlation. The rest of the procedure to obtain the monthly binned light curve is similar to those of weekly binned described above.

\subsection{Mets\"{a}hovi}
The 37~GHz observations were obtained with the 13.7 m diameter Mets\"{a}hovi radio telescope, which is a radome enclosed paraboloid antenna situated in Finland. The measurements were made with a 1~GHz-band dual beam receiver centered at 36.8~GHz. The observations are ON--ON observations, alternating the source and the sky in each feed horn. A typical integration time to obtain one flux density data point is between 1200 and 1400~s. The detection limit of our telescope at 37~GHz is on the order of 0.2~Jy under optimal conditions. Data points with a signal-to-noise ratio $< 4$ are handled as non-detections and discarded from the analysis.

The flux density scale is set by observations of the HII region DR21, with a known flux density of 17.9~Jy at 37~GHz \citep{Terasranta1998}. Sources NGC~7027, 3C~274 and 3C~84 are used as secondary calibrators. A detailed description of the data reduction and analysis is given in \citet{Terasranta1998}. The error estimate in the flux density includes the contribution from the measurement rms and the uncertainty of the absolute calibration.

\section{The Discrete Correlation Function}
\label{dcf_method}

The cross-correlation function is commonly employed in the study of AGN to probe the continuum emission mechanism by correlating multiwavelength light curves and to seek correlations between the variability and other AGN properties. The classical correlation function by \citet{Oppenheim1975} can be used for such purposes, if the data are evenly sampled. To deal with the uneven sampling, there are three variants of the classical function -- Interpolated cross-correlation function \citep{Gaskell1987}, discrete correlation function \citep[DCF; ][]{Edelson1988} and $z$-transformed DCF \citep{Alexander2013}. Due to unevenly sampled $\gamma$-ray and radio light curves of sources in this work we resorted to discrete correlation function. The DCF is estimated from the relation, 
\begin{equation}
	{\rm DCF}_{ij} = \frac{(a_i-\bar{a})(b_j-\bar{b}) }{\sigma_a \sigma_b}\,\,.
        \label{eq_dcf}
\end{equation}
where $a_i, b_j$ are the observed fluxes at times $t_i$ and $t_j$ and $\bar{a}, \bar{b}, \sigma_a$ and $\sigma_b$ are the means and standard deviation of the entire light curves. At this step, DCF$_{ij}$ are binned by their associated time lag, $\tau_{ij} = t_i - t_j$ into equal width bins. The average of the bins yields the DCF($\tau$).

The normalization (mean and standard deviation of respective light curves) used in the estimation of the DCF($\tau$) assumes that the light curves are statistically stationary. Under this assumption, the DCF sometimes can exceed unity (DCF$(\tau)>1$) making the interpretation difficult. A workaround for this problem is to estimate the mean and standard deviations only from the points that overlap at a given time lag bin \citep[c.f., ][]{White1994,Welsh1999}. Thus, with the local normalization the resultant DCF is bound to the $[-1,+1]$ interval. A positive DCF($\tau$) implies a correlated variability and an anti-correlation when its negative. The uncertainties of the DCF were estimated by a model-independent Monte Carlo method \citep{Peterson1998} that accounts for the effects of measurement noise and data sampling. The simulation consists of a bootstrap selection of a subsample of data points from each light curve to which Gaussian noise with a standard deviation matching the observational error bars are added. The time lags and 68\% fiducial interval are estimated following the maximum likelihood approach by \citet{Alexander2013}. According to this method, the maximum likelihood estimate coincides with the peak of the cross-correlation. The fiducial interval is then estimated by interpolating between the points of the likelihood function \citep[see][]{Alexander2013}.

Works such as, \citet{Welsh1999} have shown that removing a linear trend from the light curve prior to estimating the DCF to improve the estimation of time lag. However, this was found not to be true in the case of unevenly sampled light curves \citep{Peterson2004}.

\section{Significance of the correlations}
\label{dcf_sign}

The stochastic nature of the variability, data sampling and measurement noise are some of the properties that affect the result of the correlation. Also, the frequent appearance of flares means that high correlation coefficients between any two energy bands are to be expected even in the absence of any physical relation between the processes responsible for their production. Hence, we investigated the statistical significance of the DCF by performing Monte Carlo simulations, in turn, estimating the probability that the observed correlation is primarily limited by chance correlations. The widely employed procedure to estimate the significance is from the cross-correlation of simulated light curves with power law power spectral densities. Recently, two additional methods have been proposed \citep{Fuhrmann2014} to quantify the significance of the cross-correlation using mixed source correlations and stacking analysis.

We adopted all the three methods in estimating the significance and have discussed the methods in detail below. Every correlation is locally normalized as some works have shown the method to be efficient also with simulated light curves \citep[e.g.,][]{Welsh1999,MaxMoerbeck2014b}.

\subsection{Correlation of simulated light curves}
\label{dcf_simulate_LC}

The AGN light curves, in general, can be modelled by red-noise power spectra showing variability at all time scales, e.g., \citet{Hufnagel1992} in the radio and optical, \citet{Lawrence1993} in the X-rays, and \citet{Abdo2010b} in $\gamma$ rays. Hence to construct the significance level of the observed correlations, we correlate simulated light curves for both radio and $\gamma$ rays assuming a simple power law model of the form: PSD $\propto f^{-\alpha}$.

\subsubsection{Estimation of the Power Spectral Density}
\label{PSD_Methods}

The underlying PSD of the light curve is estimated by fitting a simple power law to the periodogram of the observed light curve or by fitting a straight-line to the periodogram in logarithmic scale. Due to the complications arising from {\it red-noise leak} and {\it aliasing}, we estimated the PSD using a variant of PSRESP method proposed by \citet{Uttley2002}. This method involves the simulation of light curves for a set of model parameters, estimate the periodogram for each simulated light curve and average them, and determine the best-fitting PSD from a goodness of fit. We discuss the method in detail in the following sub-sections.

\noindent
\textit{Estimation of the Periodogram.} To estimate the periodogram, every light curve (\{$t_k, x(t_k)$\} for $k$ = 1,2,\ldots,$N$) is initially binned in time intervals $\Delta T_{\rm bin}$, by taking the weighted mean of all points within each bin. Missing data points in the binned light curve were linearly interpolated before convolving with an Hanning window function to reduce red-noise leakage \citep{MaxMoerbeck2014b}.

For an evenly sampled light curve of length $N$, the periodogram is given by the modulus-squared of the discrete Fourier transform \citep[DFT;][]{Press1992} of the data. The periodogram, $P(f_j)$, at a given Fourier frequency $f_j$ \footnote{$f_j$ = $j/(N\Delta T_{\rm bin})$. The zero Fourier frequency component, $f_0$ = 0, corresponds to the sum of the light curve. For even $N, f_{N/2}$ = 1/(2$\Delta T_{\rm bin}$) is the Nyquist frequency, $f_{\rm Nyq}$.}, is then given by,
\begin{equation}
	P(f_j) = \frac{2\Delta T_{\rm bin}}{\mu^2 N}\left\{Re[DFT(j)]^2 + Im[DFT(j)]^2\right\},
	\label{eq3}
\end{equation}
where $N\Delta T_{\rm bin}/\mu^2$ is the fractional rms normalization. With this normalization, the square root of the integral of the underlying PSD between two frequencies $f_1$ and $f_2$ yields the contribution to the fractional rms squared variance (i.e., $\sigma^2/\mu^2$) due to variations from the corresponding time-scales \citep{Miyamoto1991,VanDerKlis1997}. Thus, integration between  $f_1$ and $f_{\rm Nyq}$ (even) or $f_{(n-1)/2}$ (odd) yields the total rms squared variability.

The resulting periodogram is then logarithmically transformed and binned with 10 points as discussed by \citet{Papadakis1993}. This approach reduces the scatter in the periodogram. But taking logarithm also introduces a bias to the power of the periodogram which is a constant and can be removed by adding 0.25068 to obtain the final binned logarithmic periodogram \citep{Vaughan2005}. The binned logarithmic periodogram are normally distributed within each geometric mean frequency bin.

\noindent
\textit{Simulating Light Curves.} To simulate a light curve, we specify a power spectral model which we wish to test against the data (power law in our case). The normalization of the model power spectrum is a multiplicative factor which is carried through any convolution with the window function (i.e., only the power spectral shape is distorted by sampling).

Due to the finite length of the observed light curve, power from longer than observed time-scales leaks into the shorter time-scales and distorts the observed PSD. This effect, called {\it red-noise leak}, was accounted for by simulating light curves 100 times longer than the observed light curve.

On the other hand, uneven sampling of the light curve causes variations on time-scales down to the resolution of the observed light curve by {\it aliasing}. The power above $f_{\rm Nyq}$ which is aliased to frequencies below $f_{\rm Nyq}$ makes the observed periodogram distorted. This is taken into account by limiting the resolution of the simulated light curves to be $0.1\Delta T_{\rm samp}$.

By considering the above complications, we simulate $N$ light curves assuming a power law using the method proposed by \citet{Emmanoulopoulos2013} (hereafter EMP13). The widely used algorithm of \citet{Timmer1995} (hereafter TK95) for simulating light curves is appropriate for the production of Gaussian light curves. Since the distribution of most of the light curves in our sample are far from being a Gaussian, light curves simulated using the method by TK95 may not be appropriate for the establishment of confidence intervals for PSD and cross-correlation studies. The method by EMP13 involves the combination of the routine by TK95 and the iterative amplitude adjusted Fourier transform algorithm by \citet{Schreiber1996}, producing light curves possessing exactly the PSD and the probability density function (PDF) as the observed light curve. For more information on the method the reader is referred to \citet{Emmanoulopoulos2013}.

We simulated 1000 continuous light curves using the method by EMP13. Instead of generating random numbers from the best-fitting PDF of the data, which is used during the amplitude adjustment stage in EMP13, we considered the use of cumulative distribution function (CDF). This was due to the multi-modal nature of the observed data making it very difficult to fit a function to the PDF. It is not quite so simple to estimate a PDF. If one uses a histogram one needs to choose the bin width and the starting point for the first bin. If one use the kernel density estimation one needs to choose the kernel shape and bandwidth. Hence, to circumvent all the problems of obtaining a PDF, we generated the random numbers from the CDF of the data. The CDF has a simple non-parametric estimator that needs no choices to be made -- the empirical distribution function. We also confirmed that the distribution of the generated random numbers agree with the observed data by using a two-sample Kolmogorov-Smirnov (KS) test.

The simulated light curve is finally normalized to the mean and variance of the observed light curve.

\noindent
\textit{Goodness of fit.} Once a continuous light curve is simulated, it is resampled to the sampling pattern as the observed light curve. At this stage, Gaussian noise with zero mean and variance matching those of the observations are added to the resampled light curve in the case of radio light curves, while for the $\gamma$ rays Poisson noise is added according to the equation,
\begin{equation}
	LC_{\rm sim}(t_i) \sim \frac{Pois[\mu = LC_{\rm sim}(t_i)\Delta t]}{\Delta t}\;\;{\rm for}\; i = 1,\ldots,N,
	\label{eq_pois}
\end{equation}
where $Pois[\mu]$ corresponds to the Poisson random number with a mean value of $LC_{\rm sim}(t_i)\Delta t$. The resampled light curve is then rebinned, and empty bins interpolated in the same manner as for the observed light curve. The periodogram of the resulting light curve is then obtained using equation~\ref{eq3}. We thus obtained 1000 periodograms for every simulated light curve which was binned logarithmically as was implemented for the observed periodogram. The binned logarithm periodograms are then averaged to obtain the model average periodogram, $\overline{P_{\rm sim}}(f)$, and the rms spread about the mean is also calculated and taken as the error in the power at each frequency, $\Delta \overline{P_{\rm sim}}(f)$.

Having obtained the model periodogram, we now estimate a statistic defined as $\chi^2_{\rm dist}$ \citep{Uttley2002} calculated from the model and observed periodogram, $P_{\rm obs}(f)$ as follows,
\begin{equation}
        \chi^2_{\rm dist} = \sum^{f_{\rm max}}_{f=f_{\rm min}} \frac{[\overline{P_{\rm sim}}(f) - P_{\rm obs}(f)]^2}{\Delta \overline{P_{\rm sim}}(f)^2}.
	\label{eq4}
\end{equation}
Next, we determine the $\chi^2_{\rm dist}$ between the model and every simulated periodogram by replacing the $P_{\rm obs}(f)$ with $P_{{\rm sim},i}(f)$ (where $i$ = 1,2,\ldots,1000) in equation~\ref{eq4}. The goodness of fit is then given by the percentile of the simulated $\chi^2_{\rm dist}$ distribution that exceeds the observed $\chi^2_{\rm dist}$ estimate.

This whole approach from simulating light curves to estimating the goodness of fit was tested for a range of PSD values. The PSD value with the highest goodness of fit corresponds to the best-fitting PSD. Using this PSD value we simulate 1000 light curves using the same approach as discussed above, obtain their binned logarithm periodogram and estimate the PSD for each periodogram by linear least squares method. The distribution of PSD obtained was used to get the $1\sigma$ confidence interval for the best-fitting PSD.

Thus having obtained the best-fitting PSD of all the light curves in our sample, we simulate 1000 light curves using the algorithm proposed by EMP13 for both radio and $\gamma$ rays. To every simulated light curve assuming the power law properties, we added Gaussian noise matching those of the observations for the case of radio, while for the $\gamma$ rays Poisson noise was added as shown in equation~\ref{eq_pois}. The cross-correlation function of the simulated light curves were then estimated using the same approach as for the observed light curve.

This simulation is followed by estimating the 68.27\% (1$\sigma$), 95.45\% (2$\sigma$) and 99.73\% (3$\sigma$) significance levels from the distribution of the cross-correlation coefficients for each time lag bin.

\subsection{Mixed source correlations}
\label{dcf_mixed_source}

The significance levels of the cross-correlation under this context is estimated by correlating all sources with possible combinations from our sample excluding the source under study. That is, having 55 sources in the source list, we correlate every source in the radio with all the 54 $\gamma$-ray light curves, in turn, yielding 2916 correlations. The correlations are done just as for the real data. The $1, 2$ and $3\sigma$ significance levels are estimated similar to that discussed in the above section. This approach of using every source in the sample instead of simulating light curves is under the assumption that the flares exhibited by the source at different wavelengths are physically unrelated. It is similar to the method where light curves are simulated over a range of PSD under the assumption that all the sources exhibit similar variability properties, i.e., characterized by red-noise PSDs \citep[e.g.,][]{Agudo2011a,Agudo2011b,Schinzel2011}. This also assumes that the light curves are sampled in the same way, which is strictly not true in the case of our radio observations.

\subsection{Stacking the correlations}
\label{dcf_stacking_method}

Following \citet{Fuhrmann2014}, to improve the sensitivity for the detection of correlations we consider stacking or averaging the correlations obtained from the whole source sample. We also attempted to stack the correlations obtained from the light curves which were initially normalized by dividing with the mean flux density. For further details on this method, see \citet{Fuhrmann2014}.

\section{Results}
\label{results}

\subsection{Power Spectral Density}

\begin{figure*}
	\includegraphics[width=0.9\linewidth]{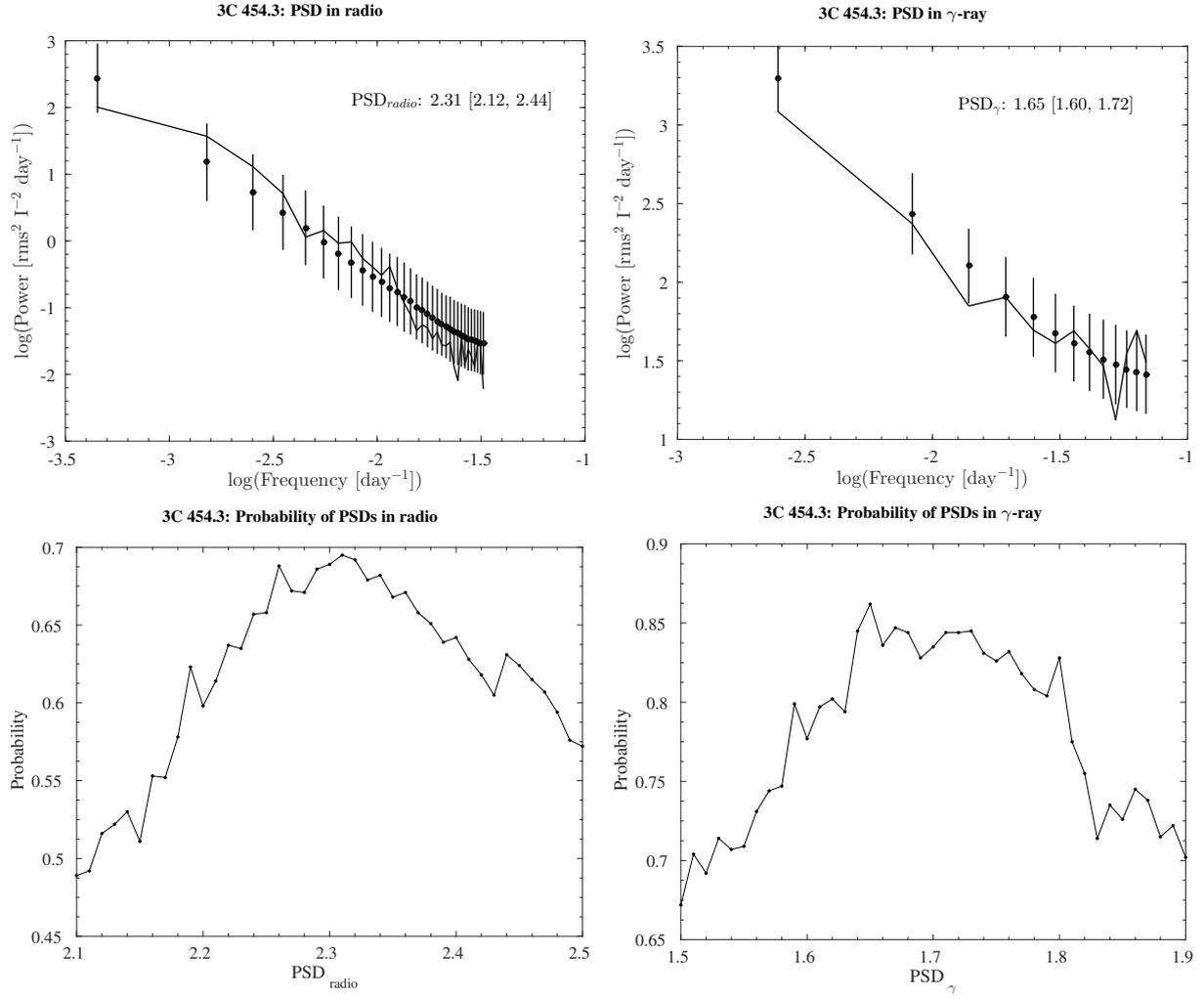}
	\caption{Power Spectral Density of the blazar 3C454.3 in $\gamma$ rays (Top left) and in radio (Top right) and their corresponding acceptance probabilities in the bottom panel. The PSD slope is shown in the figure along with its 1$\sigma$ confidence interval in square brackets.}
	\label{fig1}
\end{figure*}

\begin{figure*}
	\includegraphics[width=0.9\linewidth]{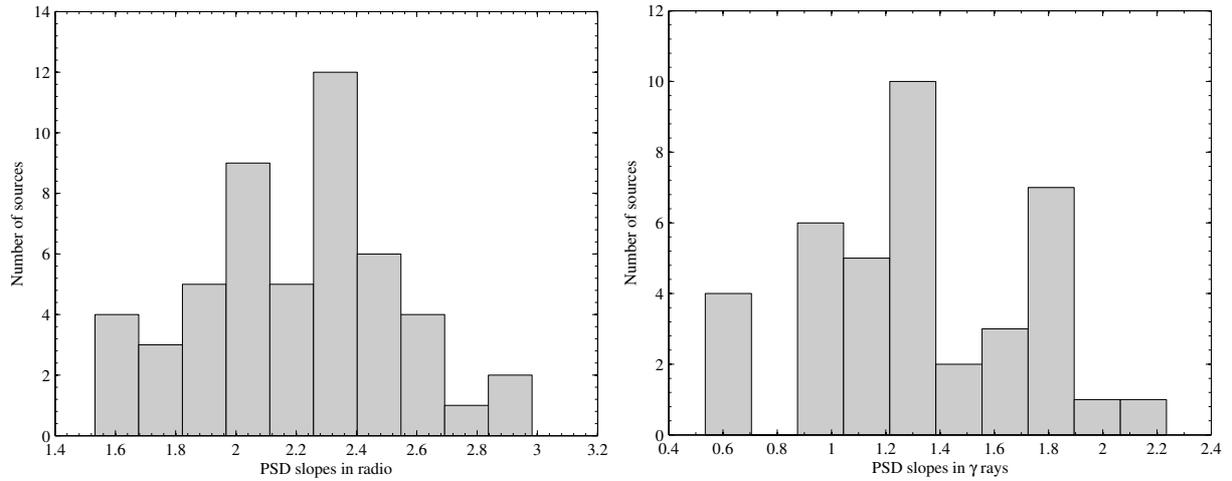}
	\caption{Distribution of PSD slopes in radio and in $\gamma$ rays.}
	\label{fig2}
\end{figure*}

The PSDs characterize the variability time-scale of a source. For estimating the PSDs of the radio light curves, we considered all the data available until 2013 August 3 to account for the long-term variability exhibited by the source. In most of the cases, we used over 20~yr of data from the Mets\"{a}hovi monitoring programme \citep{Terasranta2005}. We estimated the PSD slopes of every source at both frequencies following the Monte Carlo simulation as discussed in Section~\ref{PSD_Methods}. From the simulation we obtained the acceptance probability ($p$) which allowed us to set a constraint such that only the PSD slopes of sources with $p>$~0.05 are considered to be robust estimates. The PSD and its acceptance probability at $\gamma$ rays and radio for 3C~454.3 are shown in Fig.~\ref{fig1}. In radio we obtained the PSD slopes for 51 sources, while in $\gamma$ rays only 48 sources had acceptable estimates. The PSD slopes of other sources are taken from the average of the PSD slopes, which in radio for FSRQs and BL~Lacs are 2.2 and 2.0, respectively. In $\gamma$ rays the mean PSD slope for FSRQs and BL~Lacs are 1.3 and 1.1, respectively. This indicates that the long-term variability dominates the radio light curves, while in the $\gamma$ rays the short-term variations are more pronounced. The PSD slopes for individual sources are shown in Table~\ref{tab1}. The distribution of the PSD slopes in radio and in $\gamma$ rays are shown in Fig.~\ref{fig2}. 

Based on a comparison of our PSD estimates with earlier findings from the literature, we find:
\begin{enumerate}

	\item 3C~273: In the $\gamma$ rays, \citet{Nakagawa2013} and \cite{Sobolewska2014} reported 1.30 [1.04, 1.56] and 0.84 [0.75, 0.95], respectively, which are consistent given the errors in this work.
	\item 3C~279: In the radio, \citet{Chatterjee2008} reported a PSD slope of 2.3. Considering our higher frequency radio data and the fact that the short time-scale variability increases with frequency, our estimate of 2.00 [1.93, 2.11] is consistent with theirs.
	\item BL~Lac: \citet{Sobolewska2014} in the $\gamma$ rays obtained 0.93 [0.79, 1.11], consistent with 1.14 [1.10, 1.22] reported in this work.
	\item 3C~454.3: In the $\gamma$ rays, \citet{Nakagawa2013} obtained 1.49 [1.33, 1.65] in good agreement within the $1\sigma$ confidence interval reported in this work.
	\item \citet{Abdo2010b} using first 11 months of \textit{Fermi}/LAT data reported an averaged PSD slope for 22 FSRQs and 6 BL~Lacs to be 1.5 and 1.7, respectively. These are higher than our estimates obtained from 35 FSRQs and 12 BL~Lacs. This difference could possibly be due to the longer time range of \textit{Fermi}/LAT data considered in this work. However, the PSD slope of 1.6 [1.4, 1.8] for 3C~279 reported by the authors is consistent with our result within the quoted $1\sigma$ confidence interval.
        
\end{enumerate}

\begin{table*}
	\begin{minipage}{115mm}
	\caption{Source sample and PSD results.}
	\label{tab1}
	\begin{tabular}{cccccc}
	\toprule
	Source & 2FGL Name & Optical class & $z$ & PSD$_{\rm radio}$ & PSD$_{\gamma}$ \\
	(1) & (2) & (3) & (4) & (5) & (6) \\
	\midrule
	0059+581 & J0102.7+5827 & FSRQ & 0.644 & 1.53 [1.51, 1.61] & 1.75 [1.70, 1.85] \\
	0106+013 & J0108.6+0135 & FSRQ & 2.099 & 2.08 [2.05, 2.13] & 1.11 [1.03, 1.17] \\
	0133+476 & J0136.9+4751 & FSRQ & 0.859 & 2.31 [2.29, 2.39] & 1.05 [1.00, 1.12] \\
	0212+735 & J0217.7+7353 & FSRQ & 2.367 & 1.97 [1.84, 2.15] & 0.48 [0.43, 0.55] \\
	0215+015 & J0217.9+0143 & FSRQ & 1.721 & -- & 0.57 [0.51, 0.63] \\
	0234+285 & J0237.8+2846 & FSRQ & 1.206 & 2.52 [2.45, 2.70] & 1.65 [1.60, 1.76] \\
	0235+164 & J0238.7+1637 & BL~Lac & 0.940 & 2.23 [2.17, 2.43] & 1.15 [1.06, 1.31] \\
	0300+470 & J0303.5+4713 & BL~Lac & -- & -- & 0.63 [0.55, 0.68] \\
	3C~84 & J0319.8+4130 & GAL & 0.017 & 1.54 [1.48, 1.58] & 1.13 [1.09, 1.21] \\
	CTA~026 & J0339.4$-$0144 & FSRQ & 0.852 & 2.28 [2.22, 2.44] & 1.42 [1.39, 1.51] \\
	0420$-$014 & J0423.2$-$0120 & FSRQ & 0.916 & 1.95 [1.79, 2.00] & 1.81 [1.74, 1.86] \\
	0440$-$003 & J0442.7$-$0017 & FSRQ & 0.844 & 2.10 [2.01, 2.34] & 1.67 [1.63, 1.75] \\
	0458$-$020 & J0501.2$-$0155 & FSRQ & 2.291 & 1.72 [1.62, 1.93] & 0.84 [0.79, 0.90] \\
	0528+134 & J0530.8+1333 & FSRQ & 2.070 & 2.21 [2.13, 2.33] & 0.82 [0.78, 0.89] \\
	0605$-$085 & J0608.0$-$0836 & FSRQ & 0.870 & 2.23 [1.99, 2.31] & 1.07 [1.02, 1.14] \\
	0716+714 & J0721.9+7120 & BL~Lac & 0.310 & 2.00 [1.92, 2.15] & 0.99 [0.90, 1.06] \\
	0736+017 & J0739.2+0138 & FSRQ & 0.189 & 1.79 [1.75, 1.89] & 1.12 [1.07, 1.20] \\
	0754+100 & J0757.1+0957 & BL~Lac & 0.266 & 1.90 [1.84, 2.06] & -- \\
	0805$-$077 & J0808.2$-$0750 & FSRQ & 1.837 & -- & 1.00 [0.96, 1.11] \\
	0814+425 & J0818.2+4223 & BL~Lac & -- & 2.54 [2.43, 2.68] & 0.84 [0.78, 0.90] \\
	OJ~248 & J0830.5+2407 & FSRQ & 0.942 & 2.50 [2.35, 2.65] & 1.52 [1.49, 1.64] \\
	0836+710 & J0841.6+7052 & FSRQ & 2.218 & 1.76 [1.71, 1.87] & 1.82 [1.77, 1.91] \\
	OJ~287 & J0854.8+2005 & BL~Lac & 0.306 & 2.14 [2.04, 2.29] & 1.12 [1.06, 1.23] \\
	0917+449 & J0920.9+4441 & FSRQ & 2.188 & 2.69 [2.50, 2.97] & 1.66 [1.60, 1.77] \\
	0954+658 & J0958.6+6533 & BL~Lac & 0.367 & 1.73 [1.64, 1.87] & -- \\
	1055+018 & J1058.4+0133 & FSRQ & 0.888 & 2.19 [2.16, 2.28] & 1.59 [1.56, 1.71] \\
	1150+497 & J1153.2+4935 & FSRQ & 0.333 & 2.56 [2.42, 2.87] & 1.84 [1.79, 1.95] \\
	1156+295 & J1159.5+2914 & FSRQ & 0.725 & 2.34 [2.31, 2.42] & 1.37 [1.33, 1.47] \\
	1222+216 & J1224.9+2122 & FSRQ & 0.434 & 1.60 [1.49, 1.68] & 1.49 [1.43, 1.58] \\
	3C~273 & J1229.1+0202 & FSRQ & 0.158 & 2.27 [2.18, 2.37] & 1.18 [1.08, 1.22] \\
	3C~279 & J1256.1$-$0547 & FSRQ & 0.536 & 2.00 [1.93, 2.11] & 1.42 [1.36, 1.53] \\
	1308+326 & J1310.6+3222 & FSRQ & 0.997 & 2.08 [2.03, 2.20] & 1.27 [1.23, 1.36] \\
	1334$-$127 & J1337.7$-$1257 & FSRQ & 0.539 & 2.33 [2.24, 2.50] & -- \\
	1502+106 & J1504.3+1029 & FSRQ & 1.839 & 2.28 [2.11, 2.41] & 1.15 [1.11, 1.28] \\
	1510$-$089 & J1512.8$-$0906 & FSRQ & 0.360 & 2.39 [2.33, 2.54] & 1.38 [1.30, 1.44] \\
	1546+027 & J1549.5+0237 & FSRQ & 0.414 & 2.84 [2.76, 3.24] & -- \\
	1633+382 & J1635.2+3810 & FSRQ & 1.813 & 2.25 [2.16, 2.39] & 1.42 [1.37, 1.53] \\
	1638+398 & J1640.7+3945 & FSRQ & 1.666 & 2.24 [1.97, 2.33] & 2.14 [2.11, 2.26] \\
	3C~345 & J1642.9+3949 & FSRQ & 0.593 & 1.81 [1.78, 1.84] & -- \\
	Mark~501 & J1653.9+3945 & BL~Lac & 0.033 & 1.44 [1.41, 1.49] & 1.76 [1.71, 1.82] \\
	1730$-$130 & J1733.1$-$1307 & FSRQ & 0.902 & 1.77 [1.75, 1.84] & 1.12 [1.07, 1.23] \\
	1739+522 & J1740.2+5212 & FSRQ & 1.379 & -- & 0.83 [0.79, 0.90] \\
	1749+096 & J1751.5+0938 & BL~Lac & 0.322 & 2.27 [2.25, 2.33] & 0.60 [0.54, 0.66] \\
	1803+784 & J1800.5+7829 & BL~Lac & 0.680 & 1.89 [1.84, 1.97] & 1.01 [0.97, 1.08] \\
	3C~371 & J1806.7+6948 & BL~Lac & 0.051 & 1.88 [1.83, 1.99] & 1.20 [1.16, 1.27] \\
	4C~56.27 & J1824.0+5650 & BL~Lac & 0.664 & 2.13 [2.07, 2.29] & 1.34 [1.31, 1.43] \\
	1828+487 & J1829.7+4846 & FSRQ & 0.692 & 2.40 [2.31, 2.50] & -- \\
	2022$-$077 & J2025.6$-$0736 & FSRQ & 1.388 & 2.35 [2.23, 2.48] & 1.36 [1.32, 1.45] \\
	BL~Lac & J2202.8+4216 & BL~Lac & 0.068 & 2.02 [1.96, 2.04] & 1.14 [1.10, 1.22] \\
	2201+171 & J2203.4+1726 & FSRQ & 1.076 & 2.50 [2.24, 2.56] & 0.90 [0.85, 0.96] \\
	3C~446 & J2225.6$-$0454 & FSRQ & 1.404 & 2.17 [2.14, 2.23] & -- \\
	2227$-$088 & J2229.7$-$0832 & FSRQ & 1.559 & 2.16 [2.10, 2.39] & 1.24 [1.19, 1.33] \\
	2230+114 & J2232.4+1143 & FSRQ & 1.037 & 2.31 [2.22, 2.51] & 0.97 [0.88, 1.03] \\
	2234+282 & J2236.4+2828 & BL~Lac & 0.795 & 1.88 [1.74, 1.95] & 1.69 [1.64, 1.76] \\
	3C~454.3 & J2253.9+1609 & FSRQ & 0.859 & 2.31 [2.12, 2.44] & 1.65 [1.60, 1.72] \\
	\bottomrule
	\end{tabular}
	{\footnotesize Columns are as follows: (1) Source name; (2) 2FGL name; (3) optical classification; (4) redshift \citep{Nolan2012}; (5)(6) the first value is the best-fitting PSD obtained from Monte Carlo simulations at radio and $\gamma$-ray frequencies along with the 68.27\% confidence intervals in square brackets.}
\end{minipage}
\end{table*}

\subsection{Cross-Correlation significance in individual sources}

The cross-correlation of weekly/monthly binned $\gamma$-ray and radio light curves were performed as discussed in Section~\ref{dcf_method}. Unlike the $\gamma$-ray light curves, the radio light curves of some sources show an increasing or decreasing long-term trend which results in larger cross-correlation coefficients making the interpretation difficult. Six sources in our sample -- 3C~84, 0458$-$020, 0605$-$085, 0917+449, 3C~446 and BL~Lac -- were found to exhibit such a trend. These sources were linearly detrended prior to the correlation analysis. When investigating the time lags, we only considered lags up to half of the duration of the shortest light curve, in order to avoid spurious correlations. The significance of the DCF peak is estimated by simulating light curves of known PSDs (hereafter method 1; Section~\ref{dcf_simulate_LC}) and also from mixed source correlations (hereafter method 2; Section~\ref{dcf_mixed_source}). 

Cross-correlating weekly binned light curves and using method 1 to estimate the significance, we find 23 and 10 significant correlations at the 2$\sigma$ and 3$\sigma$ levels, respectively. With method 2 we find 38 and 6 sources significant above the 2$\sigma$ and 3$\sigma$ levels, respectively. From the cross-correlation of monthly binned light curves, we obtained 23 and 13 sources significant at 2$\sigma$ and 3$\sigma$ levels, respectively, using method 1. The number of significant sources at 2$\sigma$ and 3$\sigma$ levels were found to be 30 and 6, respectively, using method 2.

\begin{figure*}
	\includegraphics[width=0.9\linewidth]{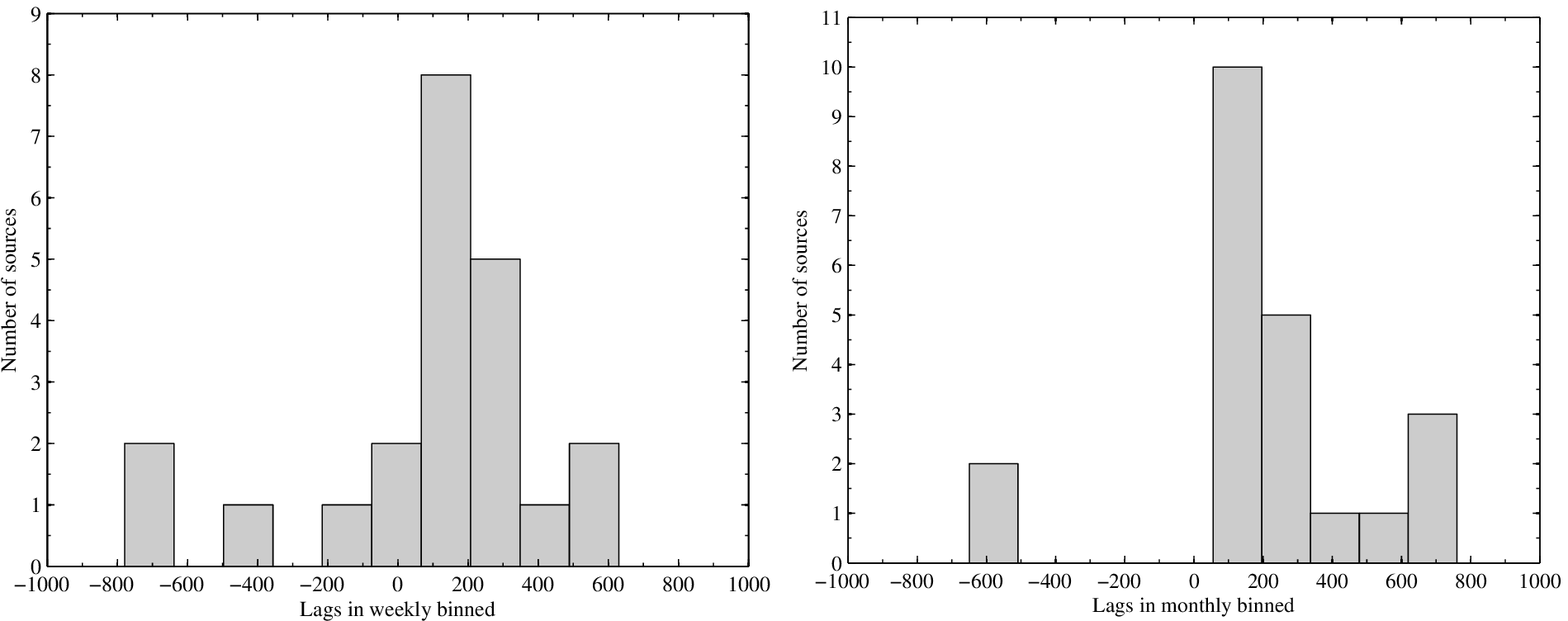}
	\caption{Distribution of Lags in weekly (Left) and monthly binned (Right).}
	\label{fig3}
\end{figure*}

Following \citet{MaxMoerbeck2014a}, we wish to set a threshold, which would produce only one spurious correlation in our sample based on chance probability. For our sample (55 sources in total), this is 98.18\% (2.36$\sigma$). We found 20 sources from the correlation of weekly binned and 23 sources from monthly binned $\gamma$-ray with radio light curves significant at $\geq2.36\sigma$ level. The time lag and the DCF peak for sources with correlation significance $>$2.36$\sigma$ level are given in Table~\ref{tab2}. A positive time lag denotes that the radio lags the $\gamma$ rays. The distribution of the lags of sources reported in Table~\ref{tab2}, using both weekly and monthly light curves, are shown in Fig.~\ref{fig3}.

We find that sources with higher significance level have the sampling of the $\gamma$-ray light curve closer to the sampling of the radio light curve, justifying the use of two binning schemes (weekly and monthly binned). However, when the significance level was the same we found no major difference in the time lags from the two binned cases. In few cases, were such a difference exists we suggest to consider the result obtained from the weekly binned $\gamma$-ray light curves.

\begin{table*}
	\begin{minipage}{175mm}
	\caption{Time lag and correlation of individual sources with significance $>$2$\sigma$. The significance level ($\sigma$) of the DCF$_{\rm peak}$ and the distance travelled by the emission region ($d_{\gamma,\rm radio}$) in parsecs are also shown. Time lags with negative sign denotes that the radio leads the $\gamma$-ray, and vice versa for time lags with positive sign.}
	\label{tab2}
	\begin{tabular}{ccccccccc}
	\toprule
	&  \multicolumn{4}{c}{Weekly} & \multicolumn{4}{c}{Monthly} \\
	\cmidrule(lr){2-5} \cmidrule(lr){6-9}
	Source & Lag & DCF & $\sigma$ & $d_{\gamma,\rm radio}$ & Lag & DCF & $\sigma$ & $d_{\gamma,\rm radio}$ \\
	& (d) &  &  & (pc) & (d) &  &  & (pc) \\
	\midrule

	0059+581 & 160 [152, 172] & 0.766 & $>$3 & 11.3 [10.7, 12.2] & 140 [130, 176] & 0.842 & $>$3 & 9.9 [9.2, 12.5] \\
	0106+013 & -- & -- & -- & -- & 350 [205, 446] & 0.640 & $>$2.36 & 44.3 [26.0, 56.6] \\
	0133+476 & 150 [142, 163] & 0.568 & $>$2.36 & 22.6 [21.4, 24.5] & 150 [120, 163] & 0.662 & $>$2.36 & 22.6 [18.1, 24.5] \\
	0215+015 & -- & -- & -- & -- & 600 [573, 720] & 0.652 & $>$2.36 & -- \\
	0234+285 & 60 [50, 96] & 0.614 & $>$2.36 & 8.5 [7, 13.6] & 90 [60, 112] & 0.651 & $>$2.36 & 12.7 [8.5, 15.8] \\
	0235+164 & 30 [16, 36] & 0.928 & $>$3 & 14.3 [7.6, 17.2] & 20 [11, 29] & 0.959 & $>$3 & 9.5 [5.2, 13.8] \\
	3C~84 & 500 [493, 508] & 0.491 & $>$2.36 & -- & -- & -- & -- & -- \\
	0420$-$014 & -- & -- & -- & -- & 20 [8, 30] & 0.615 & $>$2.36 & 1.9 [0.7, 2.8] \\
	0440$-$003 & $-$540 [$-$591, $-$505] & 0.618 & $>$2.36 & $-$20.4 [$-$22.3, $-$19.1] & -- & -- & -- & -- \\
	0805$-$077 & 120 [97, 156] & 0.590 & $>$2.36 & -- & 100 [78, 133] & 0.753 & $>$3 & -- \\
	0814+425 & 210 [128, 230] & 0.524 & $>$2.36 & -- & 200 [151, 234] & 0.703 & $>$2.36 & -- \\
	OJ~248 & $-$720 [$-$738, $-$702] & 0.829 & $>$3 & $-$88.0 [$-$90.2, $-$85.8] & $-$720 [$-$736, $-$707] & 0.916 & $>$3 & $-$88.0 [$-$89.9, $-$86.4] \\
	1156+295 & -- & -- & -- & -- & 25 [11, 92] & 0.617 & $>$2.36 & 8.6 [3.8, 31.8] \\
	1222+216 & 280 [261, 287] & 0.601 & $>$2.36 & 60.4 [56.3, 61.9] & 260 [247, 290] & 0.703 & $>$2.36 & 56.1 [53.3, 62.6] \\
	3C~273 & 160 [151, 165] & 0.625 & $>$2.36 & 29.6 [27.9, 30.6] & 160 [149, 169] & 0.770 & $>$2.36 & 29.6 [27.6, 31.3] \\
	1308+326 & -- & -- & -- & -- & 425 [412, 498] & 0.694 & $>$2.36 & 89.0 [86.2, 104.3] \\
	1502+106 & 30 [23, 88] & 0.872 & $>$3 & 2 [1.5, 5.9] & 50 [38, 63] & 0.940 & $>$3 & 3.3 [2.5, 4.2] \\
	1633+382 & $-$15 [$-$21, $-$4] & 0.686 & $>$3 & $-$2.9 [$-$4.1, $-$0.8] & 0 [$-$27, 11] & 0.701 & $>$2.36 & 0 [$-$5.3, 2.2] \\
	3C~345 & $-$40 [$-$61, $-$11] & 0.561 & $>$2.36 & $-$4.6 [$-$7.0, $-$1.3] & 30 [$-$3, 68] & 0.648 & $>$2.36 & 3.4 [0.3, 7.8] \\
	3C~345$^{a}$ & 80 [47, 101] & 0.551 & $>$2.36 & 9.2 [5.4, 11.6] & -- & -- & -- & -- \\
	1730$-$130 & 100 [75, 110] & 0.666 & $>$2.36 & 19.1 [14.3, 21.0] & 40 [31, 52] & 0.798 & $>$3 & 7.6 [5.9, 9.9] \\	
	1749+096 & 135 [120, 144] & 0.513 & $>$2.36 & 8.8 [7.8, 9.4] & 25 [14, 42] & 0.568 & $>$2.36 & 1.6 [0.9, 2.7] \\
	2022$-$077 & $-$850 [$-$861, $-$838] & 0.678 & $>$2.36 & -- & -- & -- & -- & -- \\
	BL~Lac & -- & -- & -- & -- & $-$620 [$-$641, $-$600] & 0.638 & $>$2.36 & $-$35.3 [$-$36.6, $-$34.2] \\
	2201+171 & 560 [522, 624] & 0.609 & $>$2.36 & -- & 690 [541, 703] & 0.744 & $>$2.36 & -- \\
	2230+114 & -- & -- & -- & -- & 550 [529, 576] & 0.662 & $>$2.36 & 36.1 [34.7, 37.7] \\
	3C~454.3 & 40 [32, 61] & 0.621 & $>$2.36 & 11.7 [9.3, 17.8] & 40 [30, 59] & 0.732 & $>$2.36 & 11.7 [8.7, 17.2] \\
	\bottomrule
	\end{tabular}
	\medskip
	$^{a}$ second DCF peak for 3C~345 with less significance than the first although being above the 2.36$\sigma$ level
	\end{minipage}
\end{table*}

\subsection{Stacking the correlations}

\begin{figure*}
	\centering
	\includegraphics[width=\linewidth]{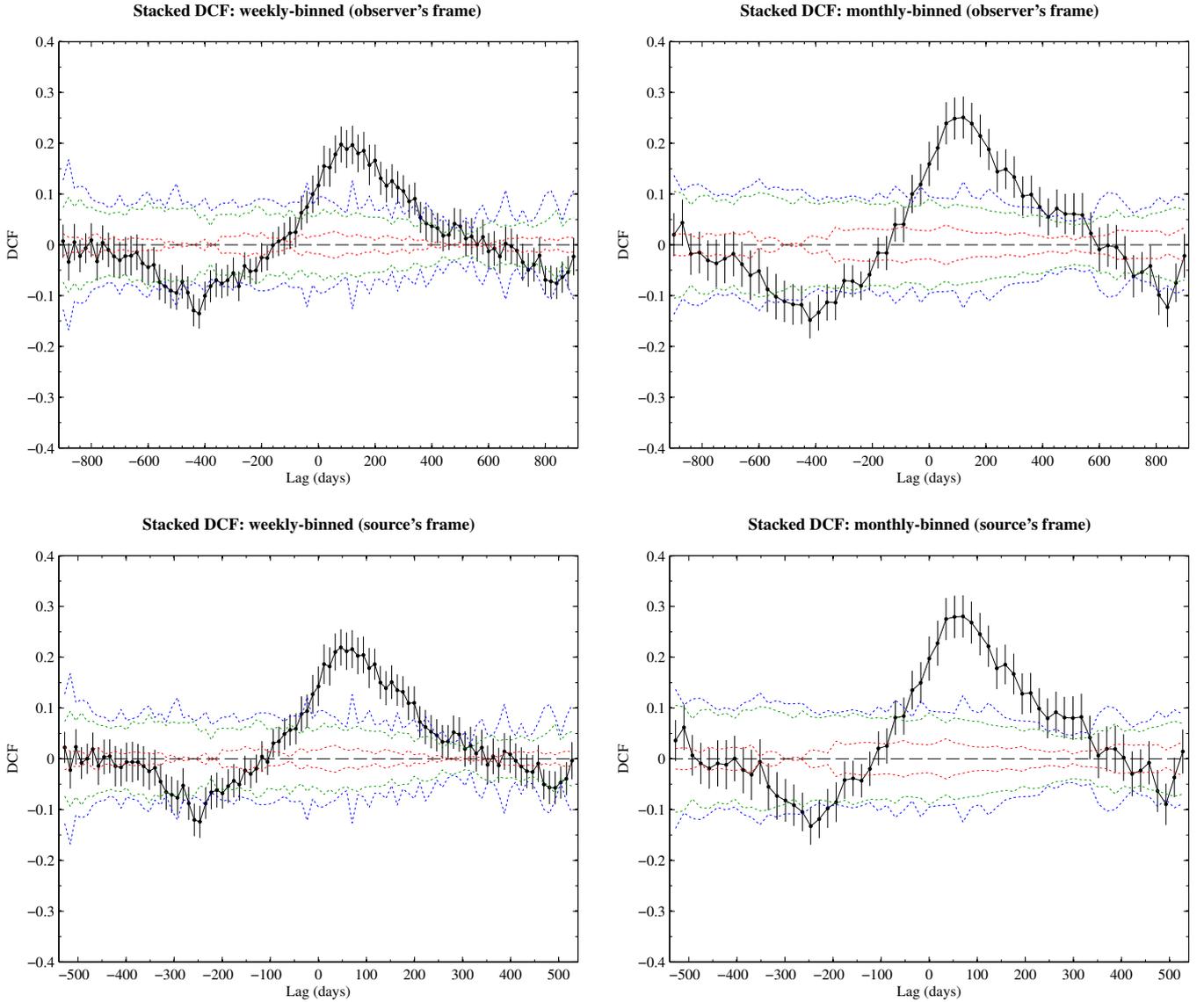}
	\caption{Stacked DCFs obtained from correlation of radio and weekly binned (Left column) and likewise from radio and monthly binned (Right column) $\gamma$-ray light curves. Stacked DCFs for the whole sample shown in observer's frame (Top) and source frame (Bottom). The red, green and blue dotted lines correspond to 1$\sigma$, 2$\sigma$ and 3$\sigma$ significance levels, respectively.}
	\label{fig4}
\end{figure*}
\begin{figure*}
	\includegraphics[width=\linewidth]{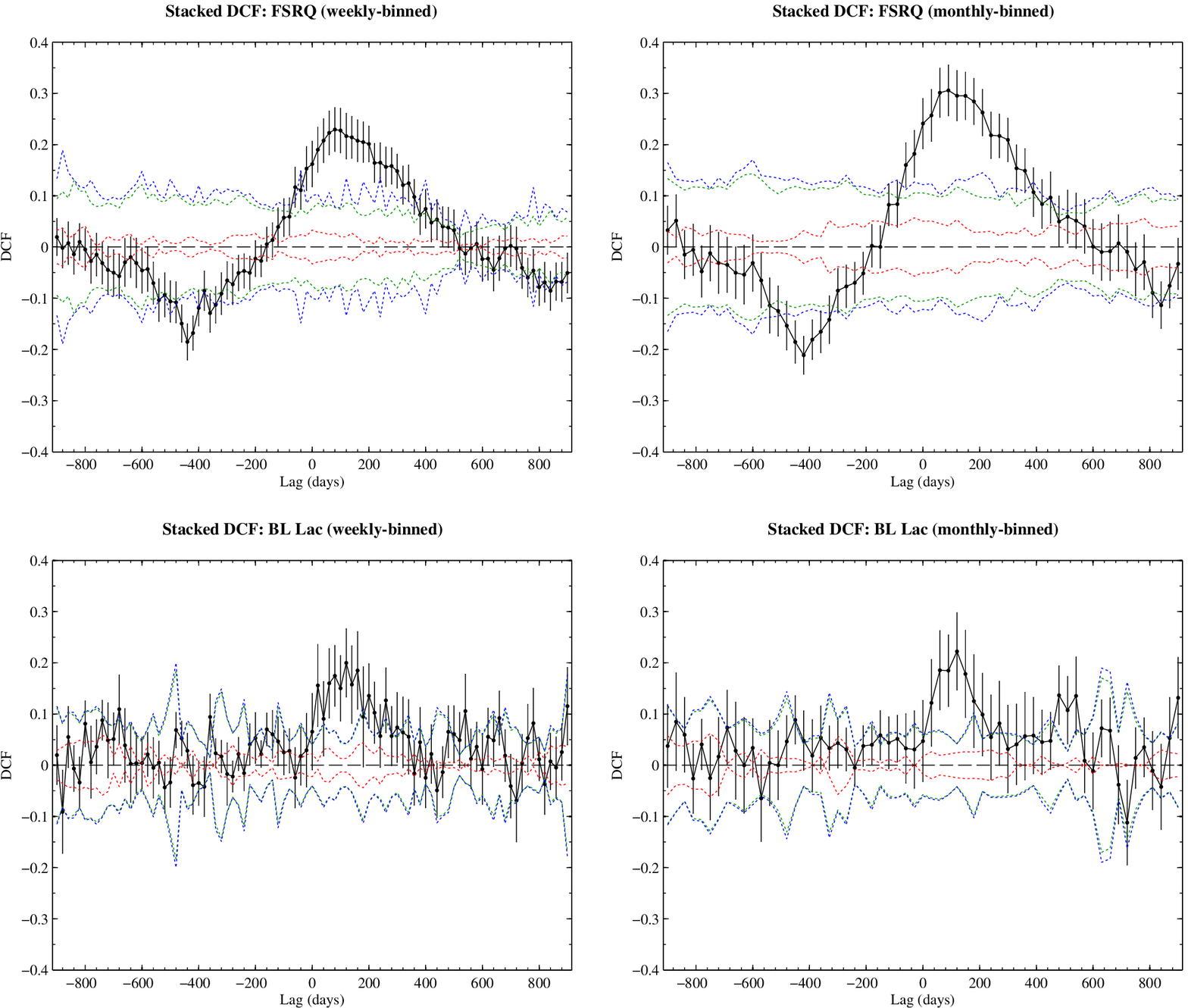}
	\caption{Stacked DCFs for the sub-sample comprising of FSRQs (Top) and for sub-sample comprising of BL~Lacs (Bottom) are shown. In left column shown are the correlation of radio and weekly binned and those from radio and monthly binned $\gamma$-ray are shown in right column.}
	\label{fig5}
\end{figure*}

The stacked DCFs were obtained following the method described in Section~\ref{dcf_stacking_method}. We got identical results from both the methods discussed. Hence, we adhered to the first method of simply averaging the correlations. We stacked the DCFs for the whole sample and also for FSRQs and BL~Lacs, respectively. The significance level for the stacked DCFs are estimated using the mixed source correlations. The stacked DCFs for the whole sample in observer's and source's frame, by scaling the time lags with a factor of 1/(1$+z$), are shown in Fig.~\ref{fig4}, while the stacked DCFs for the sub-samples are shown in Fig.~\ref{fig5}. The results obtained for the whole sample in the observer's frame are 80 and 120~d for weekly and monthly binned light curves while for those in source frame are 47 and 70~d, respectively (see Table~\ref{tab3}).

The DCFs obtained for the whole sample and for the sub-samples are significant at $>$~3$\sigma$ level. However, due to the fewer BL~Lacs in our sample, the errors of their stacked DCFs are higher. Owing to the broad DCF peak, we cannot distinguish between the time lags obtained between the weekly and monthly-binning.

To test for the possible bias on the significance of the stacked DCFs that might have been introduced by the sample selection, we removed all the sources with significance $\geq$~2.36$\sigma$. The peak of the resultant stacked DCFs was lower by a factor of $\sim$~1.2 when compared to the overall stacked DCFs but was still significant at $>$~3$\sigma$ level. This remains the same in the case of FSRQs while for the BL~Lacs the stacked DCFs were significant only at 90\%. Thus, the significance of the stacked DCFs is not affected by sources with significant correlation.

\begin{table*}
	\begin{minipage}{140mm}
		\caption{Results from stacking the correlation both in observers and source frame. Columns are similar to Table~\ref{tab2}.}
	\label{tab3}
	\begin{tabular}{cccccccc}
		\toprule
		&  & \multicolumn{3}{c}{Weekly} & \multicolumn{3}{c}{Monthly} \\
		\cmidrule(lr){3-5} \cmidrule(lr){6-8}
		Sample Type & Frame & Lag & DCF & $d_{\gamma,\rm radio}$ & Lag & DCF & $d_{\gamma,\rm radio}$ \\
			    &  & (d) &  & (pc) & (d) &  & (pc) \\
		\midrule
		\multirow{2}{*}{Whole sample} & Observer & 80 [7, 206] & 0.200 & 7.0 [0.6, 18.1] & 120 [41, 216] & 0.250 & 10.6 [3.6, 19.0] \\
					      & Source & 47 [4, 125] & 0.219 & 4 [0.3, 11] & 70 [23, 134] & 0.280 & 6 [2, 12] \\
		\multirow{2}{*}{FSRQs} & Observer & 80 [$-$1, 222] & 0.230 & 7.0 [$-$0.1, 19.6] & 90 [13, 212] & 0.306 & 7.9 [1.1, 18.6] \\
				       & Source & 47 [2, 141] & 0.229 & 4 [0.1, 12] & 53 [7, 128] & 0.306 & 4.7 [0.6, 11.3] \\
		\multirow{2}{*}{BL~Lacs} & Observer & 120 [93, 274] & 0.200 & 10.6 [8.2, 24.1] & 120 [29, 157] & 0.220 & 10.5 [2.5, 13.8] \\
					 & Source & 62 [$-$22, 83] & 0.199 & 5.5 [$-$2, 7.3] & 62 [$-$22, 106] & 0.223 & 5.5 [$-$2, 9.3] \\
		\bottomrule
	\end{tabular}
	\end{minipage}
\end{table*}

\section{Discussion}
\label{discuss}

\subsection{Correlations and comparison with earlier results}
Observations have shown that $\gamma$-ray loud AGNs are clearly associated with compact, flat radio spectrum sources \citep{Ackermann2011}. The emission and variability from the relativistic jet in the radio and the $\gamma$ rays may both be coupled to the disturbances in the central engine. The flares seen in the radio light curves are physically linked to the ejections of superluminal radio components \citep[e.g.,][]{Savolainen2002,Fromm2013}. Therefore, cross-correlation and time lags between radio and $\gamma$-ray light curves can be used to place constrains on the location of the $\gamma$-ray flares as they cannot be spatially resolved by the existing instruments.

For many sources in this work we have obtained significant correlations as shown in Section~\ref{results}. From the correlation results obtained using weekly binned $\gamma$-ray light curves, we found 3 sources (0440$-$003, OJ~248 and 2022$-$077) displaying a negative time lag (radio leading) with $\Delta t\gtrsim$~1.5~yr. Likewise, using monthly binned $\gamma$-ray light curves, two sources (OJ~248 and BL~Lac) showed a negative time lag with $\Delta t>$~1.5~yr. In all of these cases, except for BL~Lac, there is a $\gamma$-ray flare at the very beginning or end of the 5~yr period and it is difficult to judge the reliability of the correlation without complete sampling of the flares. In BL~Lac the monthly binned $\gamma$-ray light curve averages the variations too much and the weekly binning (where the correlation is not significant) is closer to the sampling of the radio light curve. Therefore, these results are not considered for further interpretation, in turn reducing the number of significant cases to 16 and 21 sources for weekly and monthly binned, respectively. For 15 sources in our sample, this is the first time a significant correlation is reported in the literature. Our results agree very well with those quoted in Table~\ref{tab4} from earlier works.

\begin{table}
        \centering
	\caption{Time lags obtained from radio/$\gamma$-ray correlation in earlier works.}
	\label{tab4}
	\begin{tabular}{cccc}
		\toprule
		Source & Lag & Frequency & Reference \\
		       & (d) & (GHz) & \\
		\midrule
		0234+285 & 40 [30, 50] & 86 & \citet{Fuhrmann2014} \\
		0235+164 & 0--50 & 15, 43 & \citet{Agudo2011b} \\
		... & $-$4 [$-$14, 6] & 86 & \citet{Fuhrmann2014} \\
		3C~273 & 120--170 & 43 & \citet{Beaklini2014} \\
		1502+106 & 14 [3, 25] & 86 & \citet{Fuhrmann2014} \\
		3C~345 & 31 [20, 60] & 43 & \citet{Schinzel2012} \\
		1730-130 & 29 [3, 55] & 86 & \citet{Fuhrmann2014} \\
		3C~454.3 & 8 [$-$4, 20] & 86 & \citet{Fuhrmann2014} \\
		... & 0 & 230 & \citet{Wehrle2012} \\
		\bottomrule
	\end{tabular}
\end{table}

\begin{figure*}
	\includegraphics[width=\linewidth]{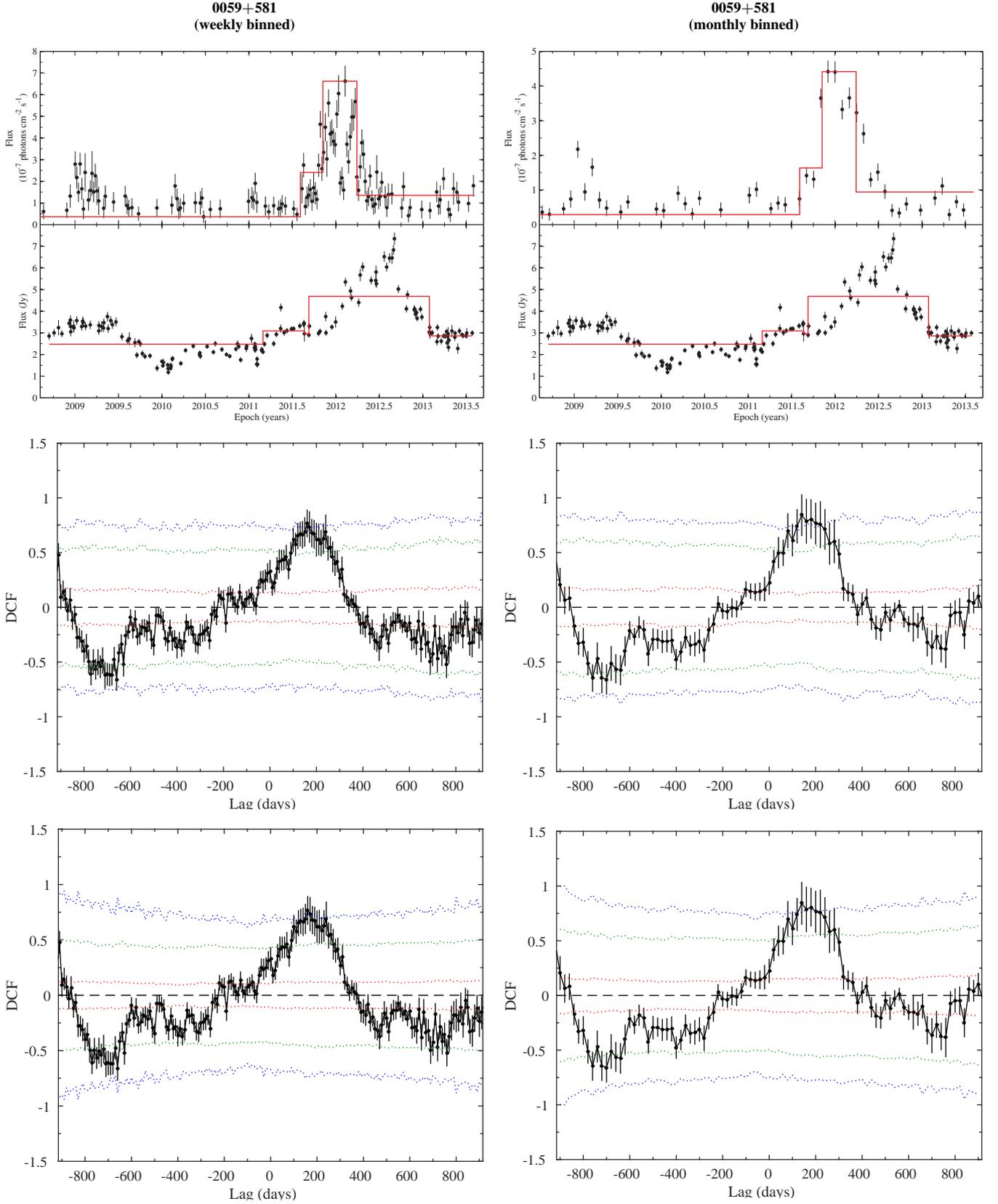}
	\caption{Light curves for 0059+581 at radio along with weekly (left column) and monthly binned (right column) $\gamma$ rays are shown in the top panel. Bayesian block representation is plotted in red line over the light curve. It is scaled relative to the light curve. The DCF for weekly and monthly binned $\gamma$-ray data with significance test using light curve simulation and mixed source method are shown in the middle and bottom panel. The 1$\sigma$, 2$\sigma$, and 3$\sigma$ significance levels are plotted in red, green, and blue dotted lines.\hspace{\textwidth}(The complete figure set is available from the online journal.)}
	\label{fig6}
\end{figure*}

\subsection{Start times of radio and $\gamma$-ray activity}
\label{ssect:start_times}

The peaks in the DCF is due to the peak-to-peak difference in the radio and $\gamma$-ray light curves. Because of larger emission regions and multiple superposed events in the radio light curves, comparing the delays between the peaks at both wavebands alone might be insufficient. Due to this issue and that owing to light-travel delay (Section~\ref{ssect:light_delay}), comparing the start times of the activity in radio and $\gamma$ rays can help us constrain the high-energy emission site and tell us more about the emission mechanisms at both wavebands \citep{AL2003,Jonathan2011}.

Hence, to estimate the beginning of the activity and characterise the variability in the sources at both wavebands, we implemented the Bayesian blocks algorithm which partitions the data into piecewise constant blocks by optimising a fitness function \citep{Scargle2013}. Using a false-positive threshold of 1\% and an iterative determination for number of blocks we obtained the Bayesian blocks representation for the light curves as shown in Fig.~\ref{fig6}. For most sources in $\gamma$ rays, we estimated the Bayesian blocks using the arrival time of photons extracted from a ROI of 1$^{\circ}$ radius centred on the source's coordinates. Due to the complex structure and variability of the $\gamma$-ray flares in the sources, 0716+714, 1222+216, 3C~273, 1510$-$089 and 3C~454.3, computing the Bayesian blocks from the photon arrival times was computationally intensive. Hence, we computed their Bayesian blocks from the respective light curves. The Bayesian blocks are shown as red line in Fig.~\ref{fig6} which in the case of $\gamma$ rays are scaled relative to the light curve shown as black circles.

With the aid of the Bayesian blocks, we can notice in most of the light curves (see Fig.~\ref{fig6}) that the start time of an outburst in $\gamma$ rays and in radio to be quasi-simultaneous. From the physical scenario, this might imply the emission at both wavebands to be associated with the same shocked feature and the seed photons responsible for the $\gamma$-ray emission might be arising either closer to or within the radio core. The former can be constrained to a region upstream of the radio core called as the acceleration and collimation zone, where the emission feature propagates along a spiral path of toroidal magnetic field peaking in the $\gamma$ rays as it exits the zone \citep{Marscher2008}. Due to synchrotron self-absorption effects, this region is opaque at radio frequencies, in which case radio flux is in a quiescent state until the moving shock interacts with the radio core. The core in the mm-wavebands has the characteristics of a standing conical shock that compresses the flow and accelerates the electrons. According to the Turbulent Extreme Multi-Zone (TEMZ) model of \citet{Marscher2014}, a standing shock oriented transverse to the jet axis at the vertex of the conical shock can create a variable nonthermal seed photon field that is highly blueshifted in the frame of the faster jet plasma, leading to rapidly variable $\gamma$-ray emission.

If the seed photons responsible for the $\gamma$-ray flare were to arise within the radio core, then the viable explanation would be the interaction of a moving shock with the core. This is shown by the correspondence of a $\gamma$-ray flare with the ejection of a superluminal component from the radio core by various VLBI analyses \citep{Jorstad2001,Schinzel2012,Jorstad2013}. The interaction of the moving shock with the quasi-stationary feature downstream of the radio core also contributes to the observed activity in the $\gamma$ rays \citep{Agudo2011a}.

\subsection{Size of the emission region}

In majority of the correlations, we found the peak of the $\gamma$-ray emission in FSRQs to precede those at radio in time-scales of days--months. Therefore, to put the correlation results into a more physical context, we estimate the distance travelled by the emission region, $d_{\gamma,\rm radio}$, from the time lags using the relation \citep{Pushkarev2010},
\begin{equation}
	d_{\gamma,\rm radio} = \frac{\beta_{\rm app}c\Delta t^{\rm obs}_{\gamma,\rm radio}}{{\rm sin}\theta (1+z)},
	\label{dist_eq1}
\end{equation}
where $\beta_{\rm app}$ is the apparent jet speed in units of speed of light $c$, $\theta$ is the jet viewing angle, $\Delta t^{\rm obs}_{\gamma,\rm radio}$ is the time lag in the observer's frame and $z$ is the redshift. We obtained $\beta_{\rm app}$ from VLBA monitoring of AGNs at 15~GHz \citep{Lister2013} and the jet viewing angle was estimated using the $\beta_{\rm app}$ and a variability Doppler factor as shown in \citet{Hovatta2009}. For 0235+164 we used the value of $\beta_{\rm app}$ from \citet{Agudo2011b}. We were unable to estimate $d_{\gamma,\rm radio}$ for six sources in Table~\ref{tab2} owing to the lack of $\beta_{\rm app}$ or variability Doppler factor. The values of $d_{\gamma,\rm radio}$ for sources having $\beta_{\rm app}$ and $\theta$ are given in Table~\ref{tab2}, along with its lower and upper limits in square brackets. We also estimated $d_{\gamma,\rm radio}$ for the time lags obtained from stacking analysis, by taking $\beta_{\rm app}$ and $\theta$ as a mean from the corresponding sample type (see Table~\ref{tab3}).

The distance, $d_{\gamma,\rm radio}$, of 7~pc estimated from the stacking analysis, corresponds to a projected distance of $\sim$~0.7~pc for an averaged redshift of 0.9 and viewing angle of 3$^{\circ}$.3 for sources with significant correlation. This corresponds to a projected size of $\sim$0.08~mas. \citet{Jorstad2001} estimated the sizes of 43~GHz VLBA core for various blazars using multi-epoch observations. They have 47 observations for the sources in our sample showing significant correlation, with the average size of 0.1~$\pm$~0.02~mas. This when compared with our estimate of 0.08~mas allows us to constrain the $\gamma$-ray emission site within the radio core, which is in line with the far-dissipation scenario in most cases.

\begin{figure*}
	\includegraphics[width=\linewidth]{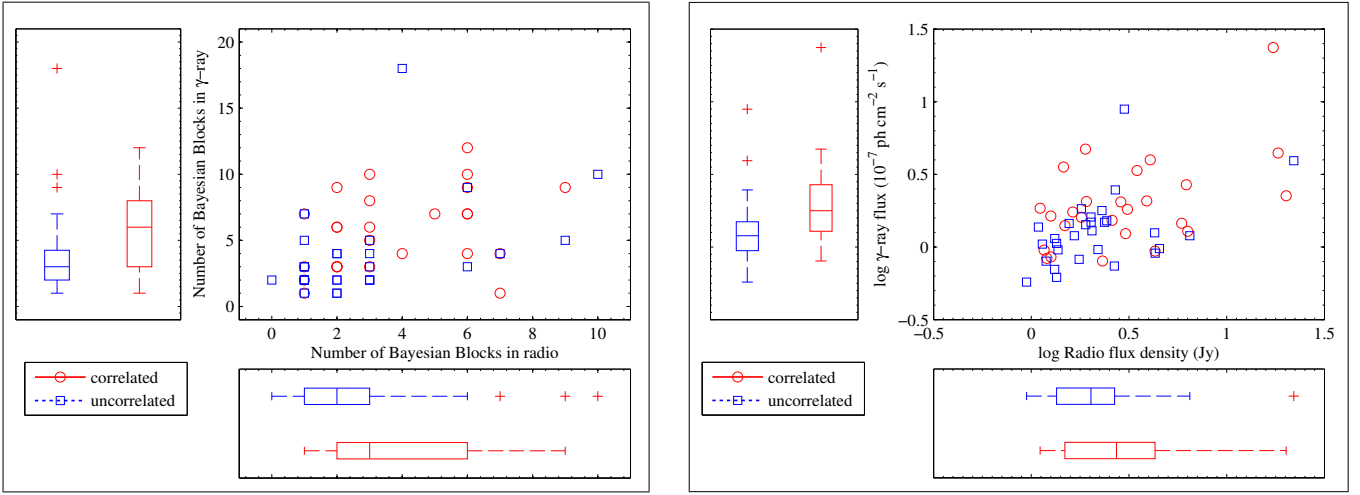}
	\caption{Scatter and boxplots for radio vs $\gamma$-rays for correlated and uncorrelated sources with number of Bayesian blocks (left) and averaged fluxes for the 5~yr duration (right). The box comprises 75\% of the distribution of data with plus symbols denoting the outliers in the boxplots. The median of the distribution is denoted by a solid line in the box. The scaling of the boxplots are similar to the scatter plot for both left-hand and right-hand panels.}
        \label{fig7}
\end{figure*}

\subsection{Light-travel argument}
\label{ssect:light_delay}

The variability in the emission of blazar jets based on the inference discussed above, is due to the shock--shock interaction over a finite size or a time interval. In either case, the radiative cooling time of electrons producing the emission in radio through synchrotron mechanism is relatively longer than the IC mechanism producing the $\gamma$ rays. Hence, the decay time-scales are significantly longer in radio than in $\gamma$ rays. Due to the size/duration of the shock interactions or the evolution of an internal shock, observations are affected by the light-travel delay so that variations faster than the light-travel time will be spatially unresolved \citep[e.g.,][]{Sokolov2004,Chen2011}.

Using light-travel delay argument, \citet{Nalewajko2014} pointed out that the temporal coincidence of radio and $\gamma$-ray flares alone cannot be used to constrain the site of the $\gamma$-ray emission, if the delay between the $\gamma$-ray and radio emission is long enough. Based on the assumption that a $\gamma$-ray flare is observed ($t_{\gamma, \rm obs}$) when a moving shock interacts with the standing shock, the authors pinpoint the time when the $\gamma$-ray photons were emitted as $t_{\gamma, \rm em} = t_{\gamma, \rm obs} - (r_{\rm core} - r_{\gamma})/c$, where $r_{\rm core}$ and $r_{\gamma}$ are the distance to the radio core and the $\gamma$-ray emission site from the central engine and $c$ is the speed of light. Thus, for the $\gamma$-ray flare to be produced within the radio core, \citet{Nalewajko2014} proposes the relation, $(r_{\rm core} - r_{\gamma}) \ll 2\Gamma^2_{\rm mm}\beta_{\rm mm}ct_{\rm mm}$ to be satisfied, where $\beta_{\rm mm}ct_{\rm mm}$ corresponds to the size of the component. Here mm corresponds to millimetre wavelength. These relations, however, hold only when a long-enough time delay is observed. We note that in many cases the $\gamma$-ray and radio activity begin simultaneously (see Section~\ref{ssect:start_times}), in which case the relation does not hold.

It should also be noted that the time-scales of variability reflect the size of the emitting region and not its distance from the central engine. The short time-scales of high-energy variability observed in many sources can still be reconciled with the emission region located parsecs away from the black hole due to very compact emission regions embedded within the jet \citep[e.g.,][]{Ghisellini2008,Giannios2009} or due to turbulence in the jet flow \citep{Marscher2014}. An alternative possibility is the formation of a small emitting nozzle in the wake of a strong recollimation \citep{Marscher2008,Nalewajko2009}.

\subsection{Uncorrelated flares}
\label{ssect:uncorrelation}

\begin{figure*}
	\includegraphics[width=\linewidth]{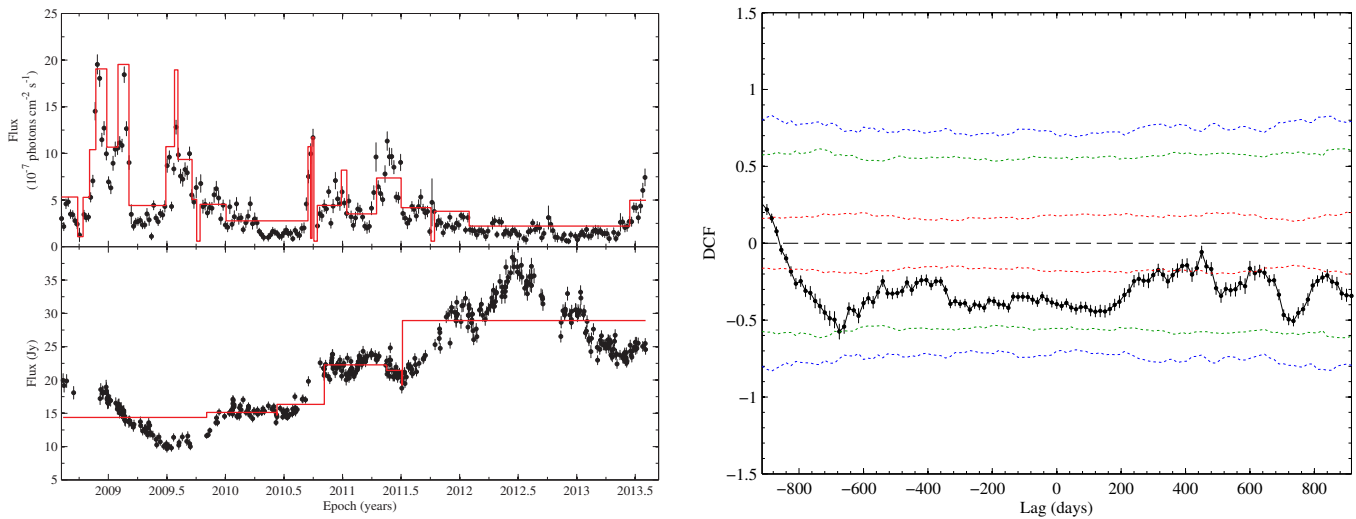}
	\caption{Left: Weekly binned $\gamma$-ray (top) and radio (bottom) light curve of 3C~279. Right: DCF of the light curve with no significant correlation.}
	\label{fig8}
\end{figure*}

In more than 50\% of the sources, we find no correlation between radio and $\gamma$ rays. To study the possible differences between the correlated and uncorrelated sources, we compared the number of Bayesian blocks (indication of flaring activity, left-hand panel of Fig.~\ref{fig7}) in the two sets of sources. Based on our false-positive threshold, we can expect at least one source to have more than one block by chance.

In the scatter plot, uncorrelated sources with two or less blocks in radio are hampered by the sampling. In the case of three blocks for uncorrelated sources in radio, we notice a major $\gamma$-ray flare with no radio counterpart or radio flare with no counterpart in $\gamma$ rays. Despite the existing uncertainty, these so-called ``orphan'' flares have been attributed to hadronic processes in the case of former \citep[e.g.,][]{Bottcher2007}, while for the latter a complex shock structure and possible shock interactions are required for producing the high-energy activity \citep[e.g.,][]{Aller2014}. According to the TEMZ model \citep{Marscher2014}, orphan flares are due to the combined effects of the modulation in the magnetic field and electron energy distribution across different turbulent cells and light-travel delay. In uncorrelated sources where four or more blocks are shown in the radio band, the lack of correlation is most likely due to the rapid variability in the $\gamma$-ray band. This, especially, applies to the sources, 0716+714, 1510$-$089 and 3C~279, comprising [10,10], [6,9] and [4,18] ([radio,$\gamma$-ray]) Bayesian blocks, which in turn, might imply the existence of different IC mechanisms for the $\gamma$-ray flares and/or to the presence of multiple emission features which manifests as smooth increase in the radio light curves due to the superposition of various events, while exhibiting multiple flares in $\gamma$ rays. As an example for an uncorrelated source, the light curve of 3C~279 along with its DCF is shown in Fig.~\ref{fig8}. We performed a two-dimensional two-sample KS-test \citep{Peacock1983}, to the estimated number of Bayesian blocks in radio and $\gamma$ rays for the correlated with the uncorrelated sources. The probability that the distributions of correlated and uncorrelated sources come from the same population is less than 4\%.

We also compared the averaged flux for the 5~yr period at both wavebands for both correlated and uncorrelated sources. This is shown using the scatter and boxplots in logarithmic scale in the right-hand panel of Fig.~\ref{fig7}. The averaged fluxes show that the correlation is significant for brighter sources at both wavebands. We also quantify this statement based on a two-dimensional KS-test as performed above, which shows that the distribution of fluxes in radio and $\gamma$ rays for correlated sources to be significantly different from the uncorrelated ones with chance probability of less than 3\%. These results indicate that only the strongest flares in the two bands are correlated, as was already suggested by \citet{AL2003}. It is possible that we obtain statistically significant results only for the strongest flares, and therefore we cannot draw strong conclusions based on the weak sources.

\section{Summary and Conclusion}
\label{conclude}

The correlation analysis using 5~yr of \textit{Fermi}/LAT and 37~GHz radio data for 55 LAT-detected blazars are presented in this work. The cross-correlation analysis of individual sources revealed 16 sources from weekly binned and 21 sources from monthly binned to be significant at $\geq2.36\sigma$ level. For 15 sources this is the first time a significant correlation is reported. In majority of the correlations, we find the peaks of the $\gamma$-ray emission to precede those at radio. We also stacked the correlations for the whole sample and for 40 FSRQs and 14 BL~Lacs to obtain a significant result and an average estimate for the corresponding sample type. The time lags for the weekly and monthly binned for the whole sample are 80 and 120~d, corresponding to 47 and 70~d in the source frame. There is no significant difference in the time lags of FSRQs and BL~Lacs upon comparison of weekly and monthly binned results.

The distance travelled by the emission region was calculated from the time lags obtained for the significant correlations. For the whole sample we obtained the distance between the occurrence of $\gamma$-ray flare and the peak of the radio flare to be $\sim7$~pc (de-projected). Two sources -- 1633+382 and 3C~345 -- showed a positive time lag with radio leading the $\gamma$ rays. This suggests that the $\gamma$-ray emission could come from downstream of the radio core \citep[e.g.,][]{Jonathan2011}.

Using the distance travelled by the emission region (projected distance $\sim$~0.7~pc) we obtained a size of $\sim$~0.08~mas. We compared this estimate with the average size of the radio core of 0.1~$\pm$~0.02~mas obtained from VLBA observations as presented in \citet{Jorstad2001}, thereby allowing us to constrain the $\gamma$-ray emission site to be co-spatial with the radio core.

Bayesian block analysis of the light curves at both wavebands shows that sources with significant correlations are more variable than the uncorrelated ones. In most of the cases, the blocks also shows that the start time of a $\gamma$-ray event corresponds closely to those in the radio, implying a co-spatial origin for the $\gamma$-ray and radio emission regions \citep[e.g.,][]{Agudo2011a}.

Sources with no significant correlation were compared by using the number of Bayesian blocks and the averaged fluxes. We find that sources with two or fewer blocks were affected by the sampling of the light curves, and for sources with four or more blocks, the uncorrelation to be due to rapid variability in the $\gamma$ rays. The average fluxes for the correlated sources were higher than for the uncorrelated ones, implying that only the strongest $\gamma$-ray flares are correlated with the radio events, as suggested by \citet{AL2003}.

These results are in favour of the far-dissipation scenario, suggesting that the origin of the seed photons for the high-energy emission is within the jet. The co-spatiality of the high-energy emission region with the radio core is the preferred scenario given the size of the emission region. This could imply that they are generated by the SSC mechanism, or by the EC mechanism with the seed photons coming from a sheath layer in the jet or an outflowing BLR. Modelling the SEDs using simultaneous observations are needed to probe this further.

Our correlation results are in good agreement with the recent findings of \citet{Fuhrmann2014}. Although the effect of synchrotron self-absorbed opacity is unavoidable at radio frequencies, we could still place constrains on the location of the high-energy emission.

\section*{Acknowledgements}
We thank the anonymous referee for providing valuable comments which improved the paper. We are also grateful for the support from the Academy of Finland to our AGN monitoring project (numbers 212656, 210338, 121148 and others). VR acknowledges the support from the Finnish Graduate School in Astronomy and Space Physics. TH was supported by the Academy of Finland project number 267324. We thank Kari Nilsson, Walter Max-Moerbeck, Dmitrios Emmanoulopoulos, Joni Tammi, and Tuomas Savolainen for useful discussions. The VLBA is an instrument of the National Radio Astronomy Observatory, a facility of the NSF, operated under cooperative agreement by Associated Universities, Inc. We also acknowledge the computational resources provided by the Aalto Science-IT project. This research has made use of NASA's Astrophysics Data System.

\bibliographystyle{mnras}
\bibliography{ms}

\label{lastpage}

\end{document}